\documentclass[twocolumn,superscriptaddress]{revtex4-1}

\usepackage{graphicx}
\usepackage{amsmath}
\usepackage{amssymb}
\usepackage{wasysym}

\newcommand{\ket}[1]{|#1\rangle}
\newcommand{\expectation}[1]{\langle #1\rangle}

\begin{document}

\title{Generalised Dicke non-equilibrium dynamics in trapped ions}

\author{Sam~Genway}
\affiliation{School of Physics and Astronomy, The
  University of Nottingham, Nottingham NG7 2RD, United Kingdom}

\author{Weibin~Li}
\affiliation{School of Physics and Astronomy, The
  University of Nottingham, Nottingham NG7 2RD, United Kingdom}

\author{Cenap~Ates}
\affiliation{School of Physics and Astronomy, The University of Nottingham, Nottingham NG7 2RD, United Kingdom}
\author{Benjamin~P.~Lanyon}\affiliation{Institut f\"ur Quantenoptik und Quanteninformation,\\ \"Osterreichische Akademie der Wissenschaften, Technikerstr. 21A, 6020 Innsbruck, Austria}\affiliation{Institut f\"ur Experimentalphysik, Universit\"at Innsbruck,Technikerstr. 25, 6020 Innsbruck, Austria}
\author{Igor~Lesanovsky}\affiliation{School of Physics and Astronomy, The University of Nottingham, Nottingham NG7 2RD, United Kingdom}

\begin{abstract}
We explore trapped ions as a setting to investigate non-equilibrium phases in a generalised Dicke model of dissipative spins coupled to phonon modes. We find a rich dynamical phase diagram including superradiant-like regimes, dynamical phase-coexistence and phonon-lasing behaviour. A particular  advantage of trapped ions is that these phases and transitions among them can be probed in situ through fluorescence. 
We demonstrate that the main physical insights are captured by a minimal model and consider an experimental realisation with Ca$^+$ ions trapped in a linear Paul trap.
\end{abstract}

\maketitle

The exploration and the understanding of the equilibrium and in particular the non-equilibrium behaviour~\cite{Kollath2007, Diehl2010, Tomadin2011, Kessler2010,*Kessler2012, Ates2012, Genway2012, *Hickey2013} of quantum many-body systems is of great current interest. Due to recent experimental advances trapped ions have become a very flexible platform to approach this challenging problem~\cite{Schindler2013,Schneider2012,Johanning2009}. In particular, spin systems~\cite{Porras2004, Deng2005} have been investigated with both a few~\cite{Kim2009} and several hundred ions~\cite{Kim2011,Britton2012} and a variety of phenomena such the emergence of interesting quantum phases~\cite{Porras2004,Friedenauer2008,Islam2011,Schneider2012}, the dynamical formation of defects~\cite{Campo2010, Chiara2010} and the role of frustration ~\cite{Kim2010, Bermudez2011} have been explored. This flexibility is rooted in the fact that the coherent coupling between electronic states (forming effective spin degrees of freedom) and the vibrations of an ion crystal can be precisely controlled.

Such coupling between spins and oscillator degrees of freedom in general leads to intriguing many-body effects. A paradigmatic example is seen in the Dicke model~\cite{Garraway2011} originally proposed to study superradiance in quantum electrodynamics~\cite{Dicke1954, Hepp1973, Wang1973}. The Dicke model features a continuous quantum phase transition at critical coupling between the spins and oscillator degrees of freedom. The transition connects a `normal' phase, where the oscillator is in its ground state to a `superradiant' phase with a displaced oscillator.  Due to the coupling of the oscillator to the spins, this transition becomes equally manifest in the polarisation of the spins. Dicke physics gives insights into a variety of phenomena such as quantum chaos~\cite{Emary2003a,*Emary2003,*Lambert2004} and the physics spin of glasses~\cite{Strack2011}. This versatility has ushered renewed interest in exploring both their statics and dynamics ~\cite{Dimer2007,Keeling2010,*Bhassen2012}. In particular much effort is currently put into experimentally realising Dicke systems out of equilibrium --- a recent example is the investigation of the non-equilibrium dynamics of superfluid cold atomic gases in optical cavities~\cite{Baumann2010,Baumann2011}.

\begin{figure}[!t]
\includegraphics[width=8.57cm]{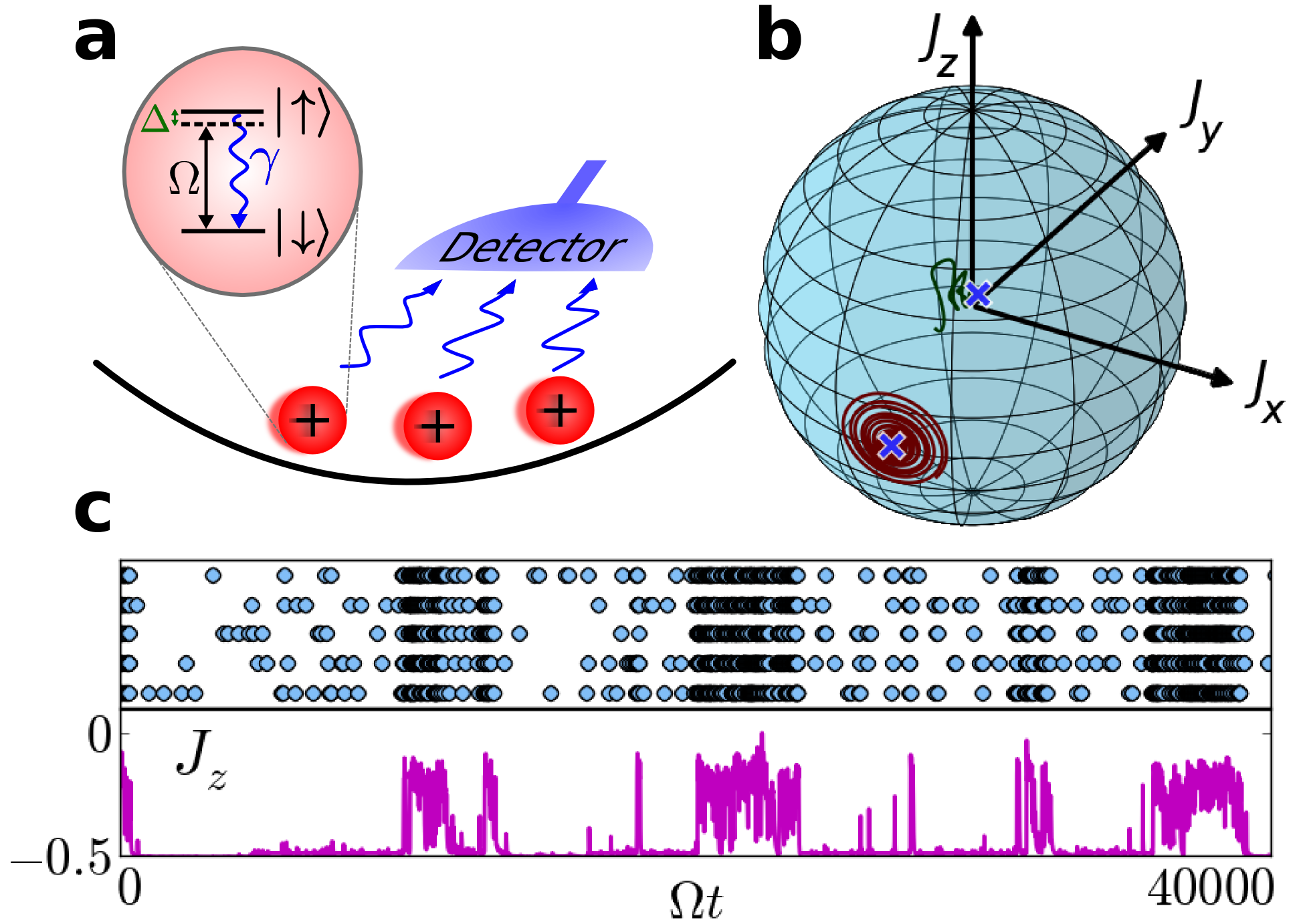}
\caption{(Colour online.) (a) Schematic diagram of a one-dimensional ion crystal.  Shown are two electronic states of each ion which represent the state $\ket{\!\!\downarrow}$ and $\ket{\!\!\uparrow}$ of a fictitious spin. Transitions between the electronic (spin) states are driven by a laser with Rabi frequency $\Omega$ and detuning $\Delta$. The state $\ket{\!\uparrow}$ relaxes to $\ket{\!\downarrow}$ at a spontaneous decay rate $\gamma$. Photons emitted in this process are detected with spatial and temporal resolution. (b) Semiclassical trajectories of the average spin polarisation on the Bloch sphere. The two fixed points ($\times$) shown correspond to two dynamical phases with strikingly different fluorescence signal, i.e. a bright and a dark state. (c) Regions in parameter space where both dynamical phases coexist are revealed by an intermittent fluorescence signal. The data shown corresponds to a Monte Carlo trajectory of 5 ions and displays the times and positions of photon emissions (top) and the spin $z$-polarisation $J_z$ (bottom). For further explanations see text.}
\label{fig1}
\end{figure}
In this work we show that a generalised version of the Dicke model --- where dissipation is introduced on the individual spins --- captures the non-equilibrium physics of a crystal of laser-driven trapped ions. The dynamical phases of this system and the associated transitions are directly linked to the time-resolved fluorescence signal from photon emissions of the ions~\cite{Ates2012}. This allows for the \emph{in situ} probing of the complex many-body out-of-equilibrium dynamics that results from a competition between coherent spin-phonon coupling~\cite{James1998,Lee2005} and tuneable incoherent spin relaxation [see Fig.~\ref{fig1}(a)]. The dynamical phase diagram of the generalised Dicke model includes non-equilibrium steady states [see Fig.~\ref{fig1}(b)] related to the traditonal `normal' and `superradiant' phases as well as a phase co-existence region. These phases become manifest in time-resolved fluorescence measurements as bright and dark regions, with phase co-existence resulting in temporal intermittency, as illustrated in Fig.~\ref{fig1}(c). In contrast with the conventional Dicke model, this generalised Dicke model has a first-order transition and the phase diagram includes a new phase where phonon lasing occurs.
 
In order to obtain a basic understanding of the physics of the driven trapped ion system, we begin by studying a minimal model of $N$ ions coupled to only a single phonon mode. Later, we generalise this to the situation found in an ion crystal formed by Ca$^+$ with many modes. The minimal model is described by the Hamiltonian
\begin{equation}
H = \Omega \sum_{i=1}^N S^x_i + \Delta \sum_{i=1}^N S_i^z + V \sum_{i=1}^N S^z_i\, (a+a^\dagger) + \omega\, a^\dag a\,.
\label{eq:H}
\end{equation}
Here, $S^x_i$, $S^z_i$ are spin operators for the internal states of the ion at position $i$ in the trap (as in Fig.~\ref{fig1}(a)) and $a$ ($a^\dagger$) is the bosonic annihilation (creation) operator for the centre-of-mass phonon mode, with frequency $\omega$.  The ions are driven with Rabi frequency $\Omega$ and detuning $\Delta$.  An electronic-state-dependent force of strength $V$ can be constructed with a far-detuned standing-wave laser field where the ions are positioned at the antinodes~\cite{Lee2005,Ripoll2005} or using amplitude-modulated laser beams~\cite{Roos2007}.  When $\Delta=0$, Eq.~\eqref{eq:H} is the Dicke Hamiltonian, with a continuous transition when $NV^2=\Omega\omega$.  Finite $\Delta$ smoothens the quantum phase transition into a crossover~\cite{Emary2004}.  

The effective decay of the spins [see Fig.~\ref{fig1}(a)] with rate $\gamma$ is captured by the dissipator
\begin{equation}
D(\rho) = \gamma\sum_i\left( S_i^- \,\rho\, S_i^+ -\frac{1}{2}\{S_i^+S_i^-,\rho\}\right)\, .
\label{eq:spindiss}
\end{equation}
The evolution of the density matrix is thus governed by
\begin{equation}
\dot\rho = \mathcal{W}(\rho) = -i[H,\rho] + D(\rho)\,.
\label{eq:W}
\end{equation}
In the following, we consider the case of trapping frequency $\omega = N\Omega$, as this scaling renders the phase boundaries $N$-independent.   
We employ a mean-field approach to study semiclassical dynamics and complement this with quantum-jump Monte Carlo simulations~\cite{Johansson2012,*Johansson2013} of~\eqref{eq:W} which illustrate what would be seen in experiment.

Using Eq.~\eqref{eq:W}, we find mean-field equations for the expectation values $A = \expectation{a}$ and the macroscopic spin polarisation $J_k = \expectation{\frac{1}{N}\sum_i S_i^k}$, for $k=x,y,z$:
\begin{align}
\dot{A} =& -\frac{i \omega}{N} A - i V J_z \phantom{\frac{X}{Y}} \nonumber \\
\dot{J_x} =&  -\frac{\gamma}{2} J_x - V J_y (A + A^*) - \Delta J_y \nonumber \\
\dot{J_y} =&  -\frac{\gamma}{2} J_y - \Omega J_z + V J_x (A+A^*) + \Delta J_x \nonumber \\
\label{eq:Z}
\dot{J_z} =&  -\gamma \left(J_z+\frac{1}{2}\right) + \Omega J_y \,.
\end{align}
\begin{figure}[tb]
\includegraphics[width=8.57cm]{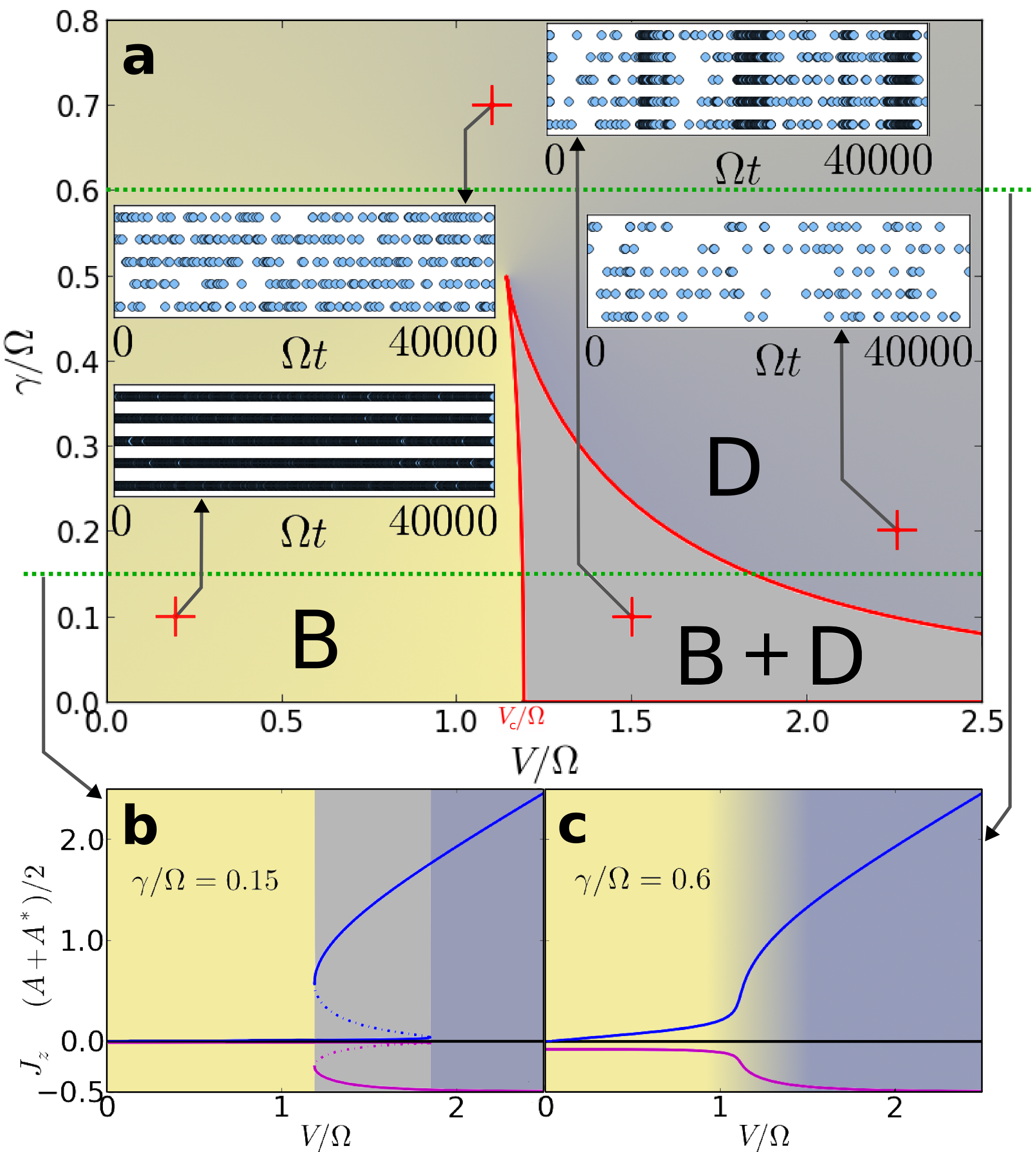}
\caption{(Colour online.) Dynamical phase diagram for $\Delta=0$. (a) The mean-field phase diagram found from the fixed-point analysis of Eqs.~\eqref{eq:Z} as a function of the coupling $V$ and inverse lifetime $\gamma$ for $\omega/N=\Omega$.  Shown are the bright (B) and dark (D) regimes and a region where phase coexistence (B+D) exists.  \emph{Inset:} quantum-jump Monte Carlo simulations, for parameter values $V$, $\gamma$ as indicated. Shown are single quantum trajectories in the steady state, with markers indicating the times at which photon emissions occur for each of the five ions along the ordinate axes. (b,c) Plots of the polarisation $J_z$ and phonon displacement $(A+A^*)/2$ at fixed points as a function of $V/\Omega$ for $\gamma/\Omega=0.15$ (b) and $\gamma/\Omega=0.6$ (c).  Solid lines indicate where the fixed points are stable. }
\label{fig2}
\end{figure}
Although information about fluctuations has been lost by replacing expectation values of products of the form $\expectation{\frac{1}{N}\sum_i S_i^k a}$ by products of expectation values $J_k A$, the equations still capture approximate average dynamics.  
We note that these equations reveal two main differences from the traditional Dicke model~\cite{Emary2003} and the Dicke model with dissipation only in the oscillator degrees of freedom~\cite{Dimer2007}.
First, the length of the macroscopic spin $J^2$ is not conserved
so that fixed points~\cite{Strogatz1994} traditionally associated with normal (zero oscillator displacement) and superradiant (large oscillator displacement) phases are different in nature.
Second, in contrast with the equilibrium Dicke model, there is no spontaneous symmetry breaking between states with opposite $J_z$ and oscillator displacement $X=(A+A^*)/2$ as the dissipation acting on the individual spins always selects the macroscopic spin-down state over a state with positive $J_z$.

Our aim is to identify steady-states of the quantum problem with the stable fixed points of the mean-field equations~\eqref{eq:Z}.
Figure~\ref{fig2}(a) shows the dynamical phase diagram found from these fixed points for $\Delta=0$ as a function of $V$ and $\gamma$. The insets show examples of time and space resolved fluorescence measurements for 5 ions obtained by quantum-jump Monte Carlo simulations of the full model~\eqref{eq:W}.  The fluorescence signals are clearly linked to the various phases. In particular, we see a  `bright phase' (B) corresponding to the state with vanishing oscillator displacement $X$ and a `dark phase' (D) when $X > 0$. In the bright phase, the effects of driving and dissipation dominate and, in our mean-field analysis, the stable fixed point lies far within the Bloch sphere.  In the dark phase, the coupling of spins to the phonon mode leads to an oscillator displacement and effective detuning which prevents the up-state from being populated.  Collective spin behaviour emerges and the mean-field fixed point lies close to the spin-down pole of the Bloch sphere [cf.~Fig.~\ref{fig1}(b)].  In the limit $\gamma\longrightarrow 0^+$ we find a simple expression for the critical coupling, $V_c$, above which the D fixed point is stable: $V_c = \left(2^{1/2}\, \Omega \omega/N\right)^{1/2}$, for general trapping frequency $\omega$.  This value differs from the non-dissipative Dicke model as does the nature of the transition itself.  It is first-order and associated with a region of phase coexistence (B+D), which occurs at finite $\gamma\apprle 0.5\Omega$, beyond $V_c$. In the fluorescence records this phase coexistence shows up as a pronounced intermittency characterised by a switching between bright and dark periods. When crossing the transition, there is a sharp change in $J_z$ associated with the emergence of a new stable fixed point of the mean-field equations [see Fig~\ref{fig2}(b)].  At large $\gamma \sim \Omega$, as shown in Fig.~\ref{fig2}(c), the transition becomes a crossover where the fixed-point moves continuously from light to dark phases with increasing $V$.  
We note that intermittency in fluorescence signals is not seen in this crossover region.

\begin{figure}[tb]
\includegraphics[width=8.57cm]{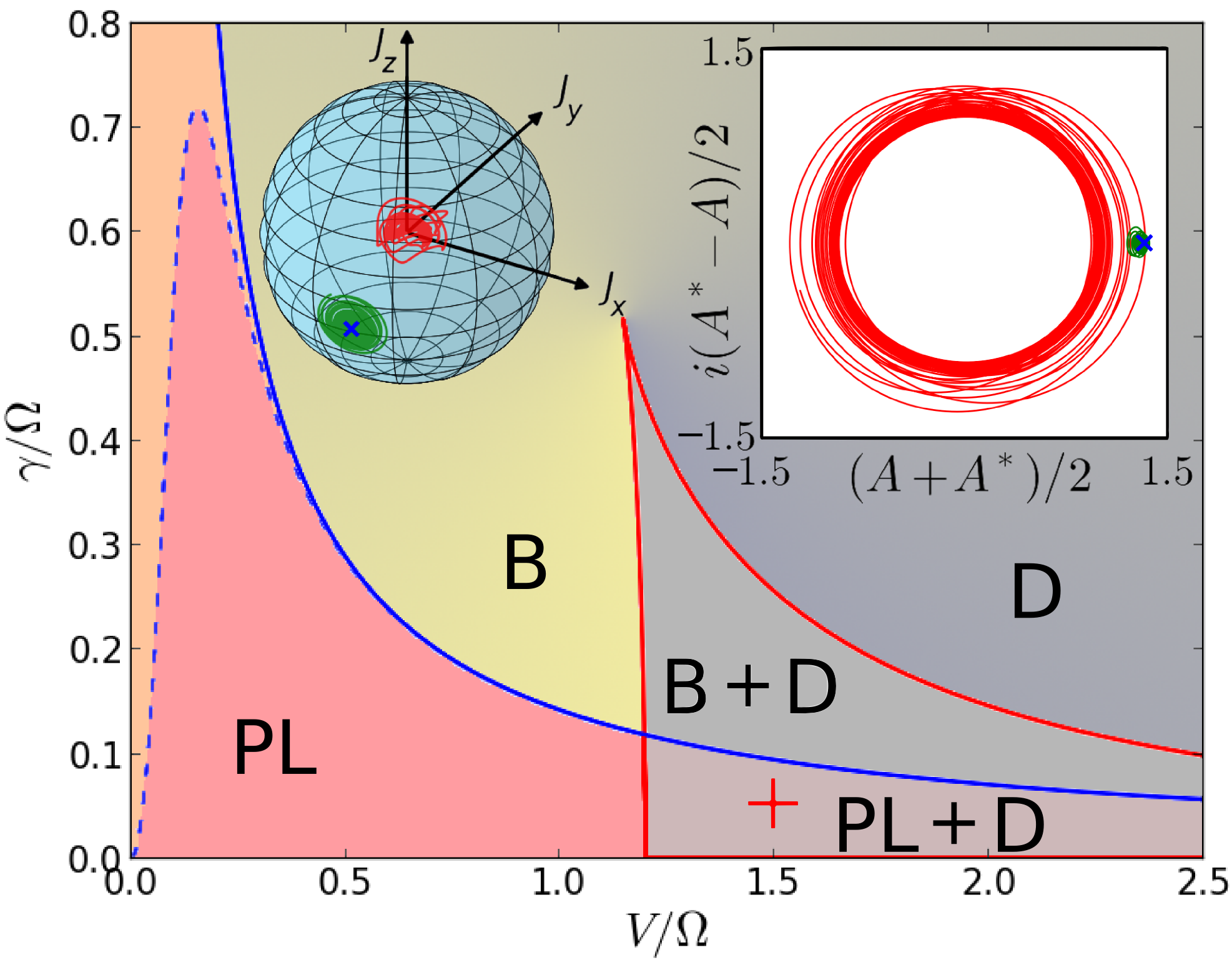}
\caption{(Colour online.) Dynamical phase diagram for $\Delta = -0.01\Omega$ with all other parameters as in Fig.~\ref{fig2}.  In addition to the bright (B), dark (D) and coexistence (B+D) regimes, points where the bright fixed point disappears at a Hopf bifurcation are shown for the case where $\kappa=0$ (solid blue line) and $\kappa/N = 10^{-4}\Omega$ (dashed blue line).  Beyond these birfurcations, limit-cycle solutions corresponding to phonon lasing (PL) exist.  Beyond the critical coupling there exists a new region (PL+D) where the dark fixed point is stable but the bright fixed point bifurcates.  The parameters indicated by `$+$' are used for the insets.  Shown inset are plots of the mean-field dynamics with different initial conditions showing the Bloch sphere and oscillator phase plane for limit-cycle (red) and fixed-point (green) steady states.  The dark fixed point is labelled `$\times$'.}
\label{fig3}
\end{figure}

Surprisingly, in contrast with the closed-system Dicke model~\cite{Emary2004}, a finite detuning $\Delta$ does not destroy the transitions shown in Fig.~\ref{fig3}.  For small $\Delta>0$,  mean-field steady states remain qualitatively unchanged and quantitatively very similar.  One might imagine that a negative detuning would compete with spin dissipation to take the system towards a steady state with a large and positive $J_z$ at large $V$, as in the closed system.  However, as shown in Fig.~\ref{fig3}, for $\Delta<0$ we find that the fixed-points at small $\gamma$ and $V$ actually become unstable at Hopf bifurcations~\cite{Strogatz1994} and, in these regimes, limit-cycle oscillations in the phonon dynamics are found at long times.  This phenomenon has been observed in trapped ions and is known as phonon lasing~\cite{Vahala2009}.  These regions, labelled PL, are shown in Fig.~\ref{fig3}, where we also plot two example semiclassical trajectories towards the dark fixed point and the limitcycle.   While it appears unphysical that the limitcycles persist down to $V=0$, this feature is destroyed by any finite dissipation on the phonons.  Such dissipation inevitably occurs in trapped-ion experiments.   We introduce a finite dissipation rate $\kappa/N \ll \gamma$ on the motional degrees of freedom by replacing $\omega \rightarrow \omega - i\kappa$ in the mean-field equations~\eqref{eq:Z}.  Changes to all dynamical phase boundaries by this dissipation are negligible, except for the phonon lasing regime, which becomes suppressed at small $V$ with finite $\kappa$ (see Fig.~\ref{fig3}).

So far we have have only included the centre-of-mass phonon mode in our considerations.  In a trapped-ion crystal, an electronic-state-dependent force on the ions constructed with a far-detuned standing-wave laser field will couple the electronic states to \emph{all} of the phonon modes.  We explore the role of the higher-frequency phonon modes by extending the simple model~\eqref{eq:H}.  The phonon modes have frequencies $\omega_m$, for $m = 1$ to $N$, and couple to the spins via the coupling Hamiltonian $\sum_{im}V_{im} S^z_i (a_m+a^\dagger_m)$.  Here $a^\dagger_m$ is the creation operator for mode $m$.  The coupling matrix elements $V_{im}$ are proportional to the normal-mode displacement vectors of the ions~\cite{James1998}, so that all vibrational modes with frequencies above the fundamental frequency $\omega_1=\omega$, have a heterogeneous phonon-spin coupling.  The dynamics of the ions are position dependent so that different spins and phonon modes have different mean-field equations of motion and Eqs.~\eqref{eq:Z} are replaced with $4N$ equations.

\begin{figure}[ht]
\includegraphics[width=7cm]{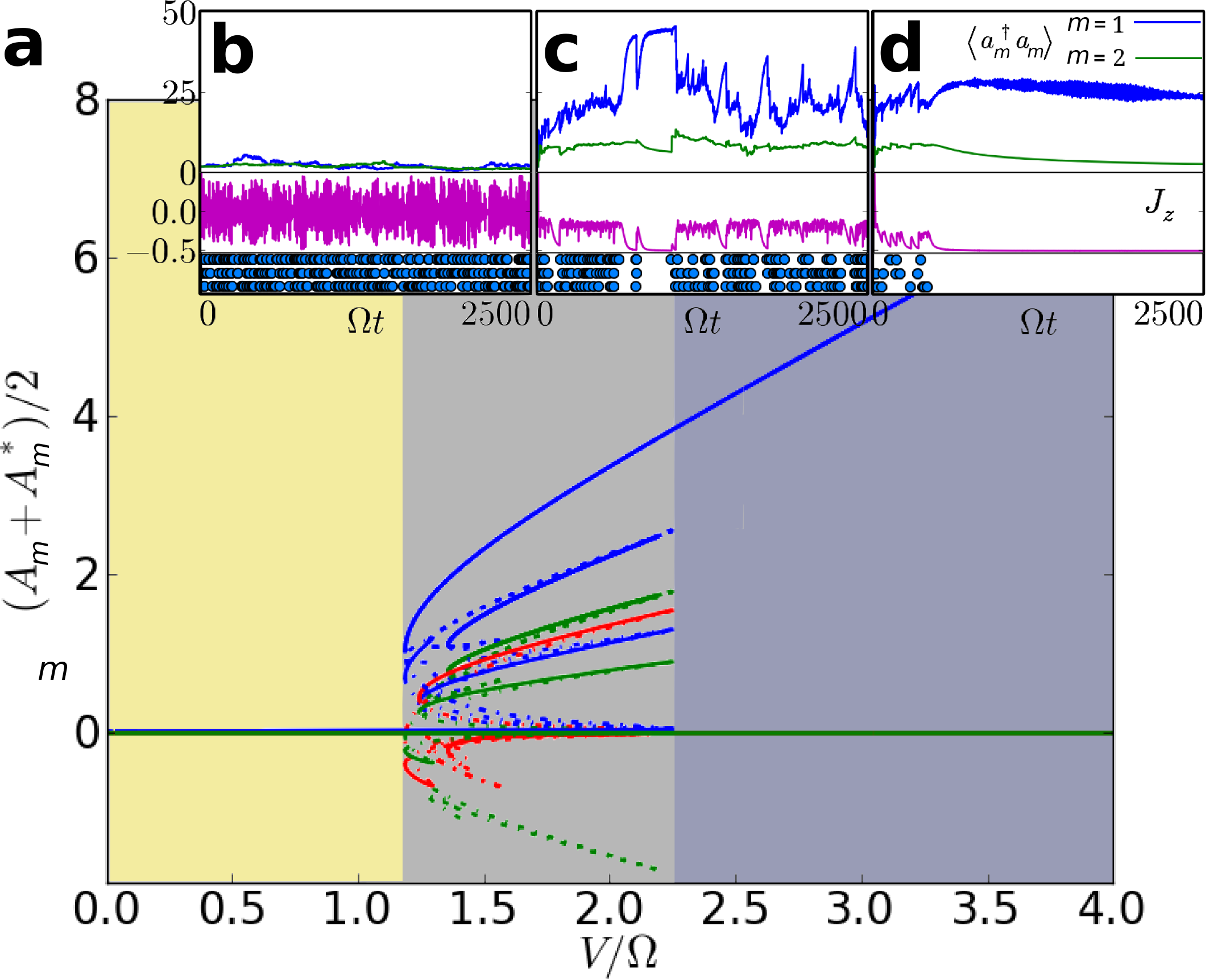}
\caption{(Colour online.) (a) Oscillator displacements at the fixed points of the three-spin three-oscillator model as a function of the coupling $V$, for $\gamma=0.1$ and $\kappa=\Delta=0$.  Regions where the fixed points are stable are shown with solid lines with all other fixed-point classifications plotted with dashed lines.  Plotted are the centre-of-mass mode with frequency $\omega_1$ (blue), and the modes~\cite{Supp} with frequencies $\omega_2$ (green) and $\omega_3$ (red). (b,c,d) Sample trajectories with phonon modes $m=1$ and 2 included showing the occupation of the two included phonon modes (top), the polarisation $J_z$ (middle) and photon emissions (bottom) of a three-ion simulation for $V=0.1$ (b), $V=1.5$ (c) and $V=3.5$ (d).}
\label{fig4}
\end{figure}

To be specific, let us study in detail the case of a three-ion system and analyse the fixed points of the 12 coupled equations associated with this three-spin, three-oscillator model~\cite{Supp}.  We find the stable fixed points in the single-mode model~\eqref{eq:Z} are also stable fixed points of many-mode dynamics~\cite{Supp}.  In addition to these steady states, in the coexistence region we find a family of additional fixed points which we illustrate in Fig.~\ref{fig4}(a).  None of these additional steady states exists at large $V \apprge 2.3\Omega$.
In Fig.~\ref{fig4}(b,c,d) we confirm these inferences by performing quantum-jump Monte Carlo simulations for three-trapped ions, including both the centre-of-mass mode and, additionally, the next mode with frequency $\omega_2$.
Although the additional phonon mode we include becomes populated close to the Dicke transition, we still observe temporal intermittency in the photon emission rate, associated with a switching between states with different $J_z$.  Crucially, far into the bright and dark regimes, our prediction that higher-frequency modes will not play a role is confirmed by simulations of quantum trajectories.

The persistence of the dark phase at strong coupling in the presence of other phonon modes can be understood qualitatively by considering the collective alignment of the spins in the down state.  We note that $\expectation{S_i^z} \rightarrow -\frac{1}{2}$, for all spins $i$, in the dark phase at large $V$.  As the spin-down state is annihilated by the Lindblad jump operator, it is stable against fluctuations arising from photon emission which disrupt collaborative behaviour.  All phonon modes except the centre-of-mass mode couple only to non-collective spin states.  As $V$ is increased, the spins become collectively polarised towards their down states, making quantum jumps less likely and the opportunities for populating other phonon modes increasingly rare.

We now check that the ion recoil due to the flourescence which allows to probe dynamical phases does not itself significantly change the dynamics.
When ion recoil is included in the model, the full Liouvillian for the spontaneous emission process couples the spatial and internal ion degrees of freedom.  For a single phonon mode, the Lindblad operators $L_i(x) = \sqrt{\gamma W(x)} e^{i\eta(a+a^\dagger)x} S_i^-$ form a continuum with $-1\le x\le 1$. $W(x) = \frac{3}{4}(1+x^2)$ reflects the angular distribution for a dipole transition and $\eta$ is the Lamb-Dicke parameter~\cite{Cirac1992}.
Repeating our fluorescence-signal simulations in Fig.~\ref{fig2} with the full Liouvillian associated with spontaneous emission
we find the collective quantum-jump behaviour qualitatively the same with  $\eta=\frac{1}{10}$.  Furthermore, in this Lamb-Dicke regime, if we expand the Liouvillian up to second order in $\eta$ as in Ref.~\cite{Cirac1992}, the semiclassical Eqs.~\eqref{eq:Z} are unchanged.

\begin{figure}[ht]
\includegraphics[width=6cm]{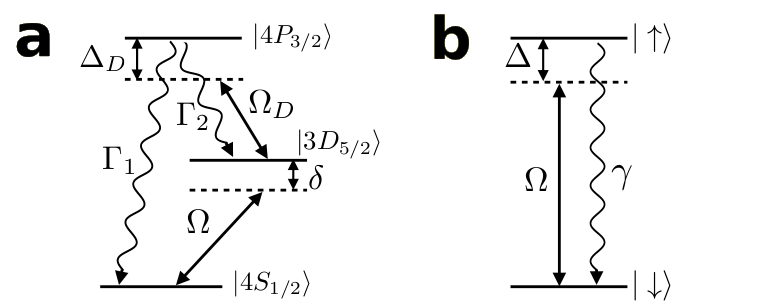}
\caption{(Colour online.) (a) Level stucture of Ca$^+$ ions driven by dressing laser of Rabi frequency $\Omega_D$ with detuning $\Delta_D$, with decay rates $\Gamma_1$ and $\Gamma_2$ shown.  The driven transition has $\Omega, \delta \ll \Omega_D \ll \Delta_D,\Gamma_1,\Gamma_2$. (b) The effective two-level scheme after projecting out the fast degrees of freedom, with unmodified Rabi frequency $\Omega$, and effective detuning and decay rate $\Delta$ and $\gamma$.  The scheme provides the states $\ket{\!\downarrow}$ and $\ket{\!\uparrow}$ which derive from $\ket{4S_{1/2}}$ and $\ket{3D_{5/3}}$ dressed with $\ket{4P_{3/2}}$.}
\label{fig5}
\end{figure}

Finally, let us show how to control the decay rate $\gamma$ which is essential for mapping out the phase diagram.  This can be tuned in a Ca$^+$-ion crystal using a simple dressing scheme.  Effective two-level systems (TLS) can be created from the ions' $\ket{4S_{1/2}}$, $\ket{4P_{3/2}}$ and $\ket{3D_{5/2}}$ states (see Fig.~\ref{fig5}).  We employ a strong dressing laser, with Rabi frequency $\Omega_D$ and detuning $\Delta_D$, between the $\ket{3D_{5/2}}$ and $\ket{4P_{3/2}}$ states. 
Projecting out the fast dynamics associated with the state $\ket{4P_{3/2}}$ using the method of adiabatic elimination~\cite{Gardiner1985}, we find effective driven and dissipative TLS with an effective dissipation rate $\gamma$ which can be adjusted by tuning $\Omega_D$.  The effective detuning $\Delta$ can be removed completely by an appropriate detuning $\delta$ of the driving laser with Rabi frequency $\Omega$.  We find effective TLS parameters $\gamma = [(\Gamma_1+\Gamma_2)\Omega_D^2]/[(\Gamma_1+\Gamma_2)^2+4\Delta_D^2]$ and $
\Delta = \delta - \Delta_D\gamma/[(\Gamma_1+\Gamma_2)]$.  For Ca$^+$, $\Gamma_1^{-1}=7.4\times 10^{-9}$s and $\Gamma_2^{-1}=101\times 10^{-9}$s so that for $\Omega_D=10$MHz and $\Delta_D=300$MHz with $\Omega=0.2$MHz and $\delta=79$kHz we have an effective TLS with $\gamma/\Omega\simeq 0.19$ and $\Delta\simeq 0$.  The system reaches an effective TLS steady-state on short time scales $\apprle 150\mu$s.  Therefore we find that the dynamical phases discussed in this work are observable on experimental time scales.

We have presented an account of dynamical phase transitions in a trapped-ion setting.  Utilising a coupling between the ions' electronic states and their centre-of-mass phonon mode, we realise a non-equilibrium Dicke model.  We have studied the system using mean-field methods which reveal a wealth of dynamical regimes, including phase coexistence and phonon lasing states which have no counterpart in the traditional Dicke model.  We have shown how dynamical phases can be observed directly via bright and dark fluorescence signals, with fluorescence intermittency exhibited in the phase-coexistence regime.

\emph{Acknowledgements.}  We wish to thank Tobias Brandes, Andrew Armour and Clive Emary for useful discussions.  This work was supported by The Leverhulme Trust under Grant No. F/00114/BG, EPSRC and the ERA-NET CHIST-ERA (R-ION consortium).  W.L.~is supported through the Nottingham Research Fellowship by the University of Nottingham. C.A.~acknowledges funding through a Marie Curie fellowship. B.L.~acknowledges support from the Austrian Research Fund under project number P25354-N20.

\bibliographystyle{apsrev4-1}
\bibliography{dicke}

\newpage

\section*{Supplemental Material}

In this supplementary material we give details of the mean-field analysis for the many-mode Dicke model.  The Hamiltonian here is
\begin{equation}
H =  \sum_i (\Omega S^x_i +  \Delta S_i^z) + \sum_{im}V_{im} S^z_i (a_m+a^\dagger_m) + \sum_m \omega_m a^\dag_m a_m
\end{equation}
where $\omega_m$, with $m = 1$ to $N$, are the $N$-ion frequency modes.  As discussed in the main text, for all vibrational modes with frequencies above the fundamental frequency $\omega_1=\omega$, the phonon-spin coupling is heterogeneous.  The coupling matrix elements $V_{im}$ are proportional to the normal-mode displacement vectors of the ions $\mathbf{b}_m$~\cite{James1998}.  
In contrast with Eqs.~\eqref{eq:Z}, the ions are dynamically distinguishable such that each spin has a different mean-field equation of motion which depends upon its position $i$ in the trap.  There are also separate equations for the average dynamics of each phonon mode, such that Eqs.~\eqref{eq:Z} are replaced with the $4N$ equations
\begin{align}
{\dot{A_m}} =& {-(i\omega_m + \kappa) A_m - i \sum_i V_{im} \Sigma_i^z \phantom{\frac{X}{Y}}}\nonumber  \\
\dot{\Sigma_i^x} =&  -\frac{\gamma}{2} \Sigma_i^x - \sum_m V_{im} \Sigma_i^y (A_m \!+\! A_m^*) - \Delta \Sigma_i^y \nonumber \\
\dot{\Sigma_i^y} =&  -\frac{\gamma}{2} \Sigma_i^y - \Omega \Sigma_i^z + \sum_m V_{im} \Sigma_i^x (A_m\!+\!A_m^*) + \Delta \Sigma_i^x \nonumber \\
\label{eq:Z2}
\dot{\Sigma_i^z} =&  -\gamma \Big(\,\Sigma_i^z+\frac{1}{2}\,\Big) + \Omega \Sigma_i^y
\end{align}
where we have denoted $\expectation{S_i^k} = \Sigma_i^k$ and $\expectation{a_m} = A_m $.  Because the fundamental frequency corresponds to centre-of-mass motion, for $m=1$, $V_{im} = V$ is $i$ independent.  For all other $m>1$, $\sum_i V_{im} =0$ and this has profound consequences for the average dynamics:  we can see immediately that the fixed points of the single-mode model (Eqs.~\eqref{eq:Z}) are also fixed points of Eqs.~\eqref{eq:Z2}.  This can be seen by noting that if all spins $i$ have the same $\Sigma_i^k$, the higher-frequency ($m>1$) oscillator modes decouple from the spins.
In the main text we study in detail the case of a three-spin model where the normal mode vectors for the three phonon modes of frequencies $\omega_1=\omega$, $\omega_2 = \sqrt{3}\omega$ and $\omega_3=\sqrt{5.8}\omega$ are $\mathbf{b}_1 = (1,1,1)/\sqrt{3}$, $\mathbf{b}_2 = (1,0,-1)/\sqrt{2}$ and $\mathbf{b}_3 = (1,-2,1)/\sqrt{6}$ respectively~\cite{James1998}.

\end{document}